\documentclass[twocolumn,10pt,aps,prd]{revtex4-1}

\usepackage{amssymb,amsmath,amsthm}
\usepackage[pdftex]{graphicx} 
\usepackage{braket}
\usepackage[T1]{fontenc}
\usepackage[latin1]{inputenc}
\usepackage[english]{babel}

\usepackage{float}
\usepackage{braket}
\usepackage{slashed}
\usepackage{commath}
\usepackage{diagbox}
\usepackage{color}
\usepackage{bm}
\usepackage[breaklinks=true]{hyperref} 
\hypersetup{colorlinks=true,citecolor=blue,linkcolor=blue,urlcolor=blue}
\usepackage{booktabs}

\addto\captionsenglish{}

\addto\captionsenglish{}

\renewcommand{\epsilon}{\varepsilon}
\newcommand{\mb}[0]{\mathbf}
\newcommand{\mn}[0]{\mu\nu}

\newcommand{\g}{g_{\mu\nu}}

\renewcommand{\rm}[1]{\mathrm{#1}}
\newcommand{\Ai}{\mathrm{Ai}}
\newcommand{\Bi}{\mathrm{Bi}}

\begin{document}
\title[]{Gravitational Particle Creation in a Stiff Matter Dominated Universe}

\author{Juho Lankinen}
\email{jumila@utu.fi}
\affiliation{Turku Centre for Quantum Physics, Department of Physics and Astronomy, University of Turku, 20014, Finland}

\author{Iiro Vilja}
\email{vilja@utu.fi}
\affiliation{Turku Centre for Quantum Physics, Department of Physics and Astronomy, University of Turku, 20014, Finland}

\begin{abstract}
A scenario for gravitational particle creation in a stiff matter dominated flat Friedmann-Robertson-Walker universe is presented. The primary creation of scalar particles is calculated using quantum field theory in curved spacetime and it is found to be strongly dependent on the scalar mass and the expansion parameter of the universe. The particle creation is most effective for a very massive scalar field and large expansion parameter. We apply the results to cosmology and calculate an upper bound for the equilibrium temperature of the secondary particles created by the  scalar field decay.
\end{abstract}

\maketitle

\section{Introduction}
Quantum field theory in curved spacetime offers a first step to merge quantum mechanics and Einstein's general relativity into a consistent theory leaving the metric tensor unquantized \citep{Birrell_Davies}. Investigations carried out by Parker \citep{Parker:1968,Parker:1969} and Zel'dovich and Starobinsky \citep{Zeldovich_Starobinsky:1972,Zeldovich_Starobinsky:1977} showed, that due to quantum effects particles are created by the expansion of spacetime. Particle creation by the changing time-dependent spacetime is also the driving mechanism behind Hawking radiation \citep{Hawking:1975}. Although these investigations were carried out in the last century, gravitational particle creation still remains as an active area of current research \citep{Haro_Elizalde:2015,Tavakoli_Fabris:2015,Quintin:2014,Fedderke:2015,Pereira:2014}.

According to the standard picture, the universe is modeled as a homogeneous and isotropic system described by Friedmann-Robertson-Walker metric \citep{FRW}. Moreover,
the very early universe is usually supposed to be very hot and radiation dominated. However, at very early times after the Big Bang there is little or no evidence that the universe had to be radiation dominated and it is possible that it was filled with some more exotic type of matter. One of these possibilities, and an interesting one, is that the early universe was dominated by stiff matter.
The possibility of an early stiff matter era was presumably first considered by Zel'dovich when he considered the consequences of an equation of state of the form $\rho=p$, where $\rho$ is the energy density and $p$ the pressure of the fluid \citep{Zeldovich:1972}. Indeed, the stiff equation of state exhibits several interesting cosmological properties worth investigating \citep{Barrow:1978}. The presence of stiff matter in the very early universe has also implications on the abundance of relic particle species produced after the Big Bang due to the expansion and cooling of the universe \citep{Kamionkowski_Turner:1990,Dutta_Scherrer:2010} and it may help explaining baryon asymmetry and density perturbations of the right amplitude for the formation of large-scale structures in our universe \citep{Joyce_Prokopec:1998}.
 Because of these important aspects, stiff matter cosmology has received renewed attention among scientists \citep{Oliveira:2011,Chavanis:2015}.

The inflationary paradigm states that our universe underwent a rapid expansion phase in the very early stages of its evolution. It is very compelling since it is able to solve a number of problems that non-inflationary cosmology is incapable of doing. These include the horizon problem, flatness of the universe and cosmic relic problems \citep{Guth:1981,Linde:1982}. Also, the inflationary paradigm is essential in explaining the structures of observed universe \citep{PLANCK}.
Stiff matter also ties in closely with inflation, since a scalar field behaves like a stiff fluid when its energy density is dominated by its kinetic energy. 
Recently, a number of models have been presented where cosmological particle production takes place at the end of an inflatory period when the universe suffers an abrupt phase transition to the stiff matter dominated phase \citep{deHaro:2016a,deHaro:2016b,Chun:2009}. 
In these models the particle production is calculated at the transition time when the universe changes phase to the stiff matter era. This approximation however, does not take into account what really happens with the particle creation when the universe expands in the stiff matter phase.

In this paper we go beyond the existing models which deal with gravitational particle creation with stiff matter. First, we present an exactly solvable model for particle creation in a stiff matter dominated universe in a more general setting. Secondly, we calculate the energy density of these created particles taking into account also the dilution effect coming from the expansion of the universe. 

The structure of this paper is as follows. In Sec. \ref{sec:II} we present the theoretical framework needed in our analysis of gravitational particle creation. We present a model for particle creation in the stiff matter dominated universe in Sec. \ref{sec:III} and study how it is dependent on the mass and momentum of a given field. We also compute the energy densities of the created particles and the background and find an upper bound for the temperature of the particles in thermal equilibrium. Finally in Sec. \ref{sec:IV} we discuss the results.
Natural units $\hbar=c=k_B=1$ are used throughout this article and the signature of the metric is chosen with positive time component.

\section{Theoretical Framework}\label{sec:II}

\subsection{Expanding Universe}
In this article we consider a universe dominated by stiff matter. The equation of state for this ideal fluid is given by the stiff matter relation $p=\rho$, where $p$ is the pressure and $\rho$ is the energy density of the fluid. Furthermore, we restrict our study to a spatially flat Friedmann-Robertson-Walker (FRW) universe.
Assuming zero cosmological constant, the dynamics is governed by Einstein's equation
\begin{align}\label{eq:EinsteinEquation}
R_{\mn}-\frac{1}{2}R\g=-8\pi G T_{\mn},
\end{align}
where $R_{\mn}$ is the Ricci tensor, $R$ the Ricci scalar, $T_{\mn}$ the energy-momentum tensor and the gravitational constant is denoted by $G$. Furthermore, if the matter is modeled by a perfect fluid with total energy density $\rho_{\rm{tot}}$, 
the Friedmann equation 
\begin{align}\label{eq:Friedmann}
\frac{\dot{a}(t)^2}{a(t)^2}=\frac{8\pi G}{3}\rho_{\rm{tot}}(t)
\end{align}
determines the time-evolution of the scale factor $a(t)$.

In the case we are considering the total energy density is comprised of the background energy density $\rho_{\rm{stiff}}$, corresponding to the stiff matter, from the energy density $\rho_{\phi}$ of the massive scalar particles $\phi$ and the density $\rho$ of some ordinary relativistic particles. In the early stages when particle creation has not yet began, we suppose that the energy densities $\rho_\phi$ and $\rho$ can be neglected, i.e., the universe is stiff matter dominated, $\rho_{\rm{tot}}= \rho_{\rm{stiff}}$.
As the universe expands the energy of the stiff matter is converted to the scalar particles $\phi$ when $\rho_{\rm{stiff}}$ decreases until 
at some time $t_{\rm{eq}}$ the energy densities $\rho_{\phi}$ and $\rho_{\rm{stiff}}$ will become equal ending the stiff matter dominated era. Supposing that
the scalar particles decay to relativistic degrees of freedom it is possible to 
evaluate their temperature as the energy density of these relativistic particles is given by
\begin{align}\label{eq:Rho_Rel}
\rho=g_*\frac{\pi^2}{30}T^4,
\end{align}
where $g_*$ denotes the number of relativistic effective degrees of freedom. 
Using Eq.\ \eqref{eq:Rho_Rel} we can obtain an upper bound for the temperature of the particles in thermal equilibrium as
\begin{align}\label{eq:T_eq}
T_{\rm{th}}\leq \Big[\frac{30}{g_* \pi^2}\rho_\phi(t_{\rm eq})\Big]^{1/4},
\end{align}
where the equality is valid whenever the scalar particles decay and thermalize very fast. 
After the ordinary particles have thermalized, the universe has entered the conventional radiation dominated era.

\subsection{Quantum Field Theory in Curved Spacetime}
The next step is to describe the production of massive scalar particles during the stiff matter dominated era.
These particles are described as excitations of quantum fields propagating in a classical background.
In curved space the Klein-Gordon equation for a massive free scalar field $\phi$ is
\begin{align}\label{eq:KG_eq}
(\square+m^2+\xi R)\phi(t,\mb x)=0,
\end{align} 
where $m$ is the mass of the field, $\xi$ a dimensionless coupling constant, $R$ is the Ricci scalar and the covariant d'Alembert operator is denoted by $\square$. We use 
the conformal coupling, where $\xi=1/6$ in four spacetime dimensions.

The field can be decomposed in any orthonormal basis solutions $u_{\mb k}$ of the Klein-Gordon equation as
\begin{align}
\phi(t,\mb x)=\sum_{\mb k}[a_{\mb k} u_{\mb k}(t,\mb x)+a^\dagger_{\mb k} u^*_{\mb k}(t,\mb x)],
\end{align}
where the wave number $\mb k$ can be discrete or continuous. The creation and annihilation operators $a_{\mb k}^\dagger$ and $a_{\mb k}$ satisfy the commutation relations $[a_{\mb k'},a_{\mb k}^\dagger]=\delta_{\mb{kk'}}$ with all other commutators vanishing. The annihilation operator defines a vacuum state $\ket{0}$ through $a_{\mb k}\ket{0}=0,\forall\, {\mb k}$ associated to the basis $u_{\mb k}$. However, in curved space other inequivalent choices of the orthonormal basis exists \citep{Birrell_Davies} and the field $\phi$ can be decomposed also in terms of these other modes $\tilde{u}_{\mb k}$ as
\begin{align}
\phi(t,\mb x)=\sum_{\mb k}[\tilde{a}_{\mb k} \tilde{u}_{\mb k}(t,\mb x)+\tilde{a}^\dagger_{\mb k} \tilde{u}^*_{\mb k}(t,\mb x)],
\end{align}
where $\tilde{a}_{\mb k}$ and $\tilde{a}_{\mb k}^\dagger$ are the annihilation and creation operators corresponding to the modes $\tilde{u}_{\mb k}$ and $\tilde{u}_{\mb k}^*$ and an other vacuum state $\ket{\tilde{0}}$ is defined through $\tilde{a}_{\mb k}\ket{\tilde{0}}=0,\forall \,{\mb k}$.
Any two sets of modes are related by a Bogoliubov transformation
\begin{align}\label{eq:Bogoliubov1}
\tilde{u}_{\mb k}=\sum_{\mb k'}[\alpha_{\mb{kk'}}u_{\mb k'}+\beta_{\mb{kk'}}u^*_{\mb k}],
\end{align}
where the coefficients $\alpha_{\mb{kk'}}$ and $\beta_{\mb{kk'}}$ are known as Bogoliubov coefficients \cite{Birrell_Davies}. 
Furthermore, if the modes $u_{\mb k}$ are separable into space and time dependent factors, then the Bogoliubov coefficients fulfill the relations $\alpha_{\mb k \mb k'}=\alpha_k\delta_{\mb k\mb k'}$ and $\beta_{\mb k\mb k'}=\beta_k\delta_{-\mb k\mb k'}$  and the relation \eqref{eq:Bogoliubov1} can be expressed modewise as  \citep{Birrell_Davies}
\begin{align}\label{eq:Bogoliubov2}
u_{\mb k}=\alpha_k \tilde{u}_{\mb k}+\beta_k \tilde{u}_{-\mb k}^*,
\end{align}
where $k=|\mb k|$.

In any case, the creation and annihilation operators of two different
bases are related via a corresponding Bogoliubov transformation, so that the expectation value of the particle number operator $N_{\mb k}=a_{\mb k}a_{\mb k}^\dagger$ of 
the modes $u_{\mb k}$  in the vacuum of $\tilde{u}_{\mb k}$ modes is given by the Bogoliubov coefficient $\beta_{\mb{kk'}}$ as
\begin{align}
\braket{\tilde{0}|N_{\mb k}|\tilde{0}}=\sum_{\mb k'}|\beta_{\mb{kk'}}|^2.
\end{align}
Thus, two observers at rest in distinct coordinate systems (with mode expansions $u_{\mb k}$ and $\tilde{u}_{\mb k}$) would see different particle content of the universe.

Since the notion of a vacuum state in curved space is ambiguous, a question arises on how to define a physical vacuum state. In FRW universe a well-known vacuum state based on adiabatic field modes is given by the so-called adiabatic vacuum state first introduced by Parker \cite{Parker:1969}. In general finding the field modes reguires solving the Klein-Gordon equation for a given metric.
For spatially flat FRW metric the field modes can be separated into time and space factors. 
Using the conformal time $\eta$ given by $d\eta=dt/a(t)$, the modes can be written as 
$u_{\mb k}\propto e^{i \mb k\cdot \mb x}\chi_k(\eta)$ so that the Eq. \eqref{eq:KG_eq} reduces to a set of equations for time-dependent harmonic oscillators
\begin{align}\label{eq:ChiEq}
\frac{d^2}{d\eta^2}\chi_k(\eta)+\omega_k(\eta)^2\chi_k(\eta)=0,
\end{align}
where $\omega_k(\eta)^2= k^2+C(\eta)m^2$ and $C(\eta)=a(t)^2$ is the conformal scale factor. The normalization of these modes is equivalent to the Wronskian condition 
$\chi_k\partial_\eta\chi_k^*-\chi_k^*\partial_\eta \chi_k=i$. Equation \eqref{eq:ChiEq} possesses a formal solution
\begin{align}\label{eq:WKB_solution}
\chi_k(\eta)=\frac{1}{\sqrt{2W_k(\eta)}}\exp\Bigl[-i\int_{\eta_0}^{\eta} W_k(\eta')d\eta' \Bigr],
\end{align}
where the function $W_k$ satisfies a non-linear equation
\begin{align}\label{eq:NonLinear}
W_k(\eta)^2=\omega_k(\eta)^2-\frac{1}{2}\Big( \frac{\ddot{W}_k}{W_k}-\frac{3}{2}\frac{\dot{W}_k^2}{W_k^2}\Big).
\end{align}
Here the dot means derivative with respect to the conformal time. Solutions for Eq. \eqref{eq:NonLinear} can be approximated by iteration, but 
if the spacetime is slowly varying, i.e., 
\begin{align}
\frac{d^l}{d\eta^l}\frac{\dot{C}(\eta)}{C(\eta)}\overset{\eta\to\pm\infty}{\longrightarrow} 0,
\end{align} 
for all $l\geq 0$, the iterated, adiabatic modes become exact.
\section{The Model}\label{sec:III}

\subsection{Gravitational Particle Creation}
To proceed, we consider a four-dimensional spatially flat Robertson-Walker universe for which the line element is given by
\begin{align}\label{eq:RWConf}
ds^2=C(\eta)(d\eta^2-dx^2-dy^2-dz^2),
\end{align}
with the conformal factor
\begin{align}
C(\eta)=b^2\eta, \quad 0 < \eta < \infty.
\end{align} 
Here $b$ is a positive constant controlling the expansion rate of the universe. The model describes a stiff matter filled universe and the standard time scale factor $a(t)\propto t^{1/3}$ and the energy density $\rho_{\rm{stiff}}$ for the stiff matter as $\propto a(t)^{-6}$. 

By inserting the metric \eqref{eq:RWConf} into the Klein-Gordon equation \eqref{eq:KG_eq} we obtain a differential equation
\begin{align}\label{eq:airydiff}
\frac{d^2\chi_k}{d\eta^2}+(k^2+b^2m^2\eta)\chi_k=0.
\end{align}
Equation \eqref{eq:airydiff} reduces to an Airy differential equation \citep{Abramowitz_Stegun} with the solution
\begin{align}
\chi_k(\eta)=C_1 \Ai\Big( \frac{-k^2-b^2m^2\eta}{(-b^2m^2)^{2/3}} \Big)+C_2 \Bi\Big( \frac{-k^2-b^2m^2\eta}{(-b^2m^2)^{2/3}} \Big),
\end{align}
where $C_1$ and $C_2$ are constants which are determined by the Wronskian condition. The spacetime is slowly varying in the asymptotic future so that the adiabatic modes become exact. Using asymptotic formulae for the Airy functions and comparing them with Eq. \eqref{eq:WKB_solution} we can recognize that normalized positive modes in the asymptotic future are
\begin{align}
\chi_k^{\rm{out}}(\eta)=\frac{e^{i\pi/12}\sqrt{\pi}}{(bm)^{1/3}}\Ai\Big( \frac{-k^2-b^2m^2\eta}{(-b^2m^2)^{2/3}}\Big).
\end{align}
On the other hand the normalized positive modes near the initial time $\eta=0$ are recognized as
\begin{align}
\chi_k^{\rm{in}}(\eta)=\frac{1}{\sqrt{2k}}e^{-i k\eta}.
\end{align}
These two sets of modes are related by the Bogoliubov transformation \eqref{eq:Bogoliubov2} as
\begin{align}\label{eq:Bogoliubov_Chi}
\chi_k^{\rm{out}}=\alpha_k\chi_k^{\rm{in}}+\beta_k\chi_k^{\rm{in}*}.
\end{align}
Keeping the coefficients $\alpha_k$ and $\beta_k$ constant and taking the derivative with respect to the conformal time in Eq. \eqref{eq:Bogoliubov_Chi} we get a system of equations from where the Bogoliubov coefficients can be solved. By keeping the coefficients constant, we ensure that the normalization of the Bogoliubov coefficients $|\alpha_k(\eta)|^2-|\beta_k(\eta)|^2=1$ is in force at all times \citep{Zeldovich_Starobinsky:1972}. For the square of the absolute value of the coefficient $\beta_k$ we obtain
\begin{widetext}
\begin{equation}\label{eq:BetaSq}
|\beta_k|^2=\frac{\pi}{8kz^{2/3}} \Big\{\!-\!\frac{2kz^{2/3}}{\pi}+k^2\Big[\Ai\Big(\frac{-k^2-z^2\eta}{z^{4/3}}\Big)^2\!+\Bi\Big(\frac{-k^2-z^2\eta}{z^{4/3}}\Big)^2 \Big] +z^{4/3}\Big[\Ai'\Big(\frac{-k^2-z^2\eta}{z^{4/3}}\Big)^2 \!+\Bi'\Big(\frac{-k^2-z^2\eta}{z^{4/3}}\Big)^2 \Big] \Big\},
\end{equation}
\end{widetext}
where $z= mb$ and $\Ai'$ and $\Bi'$ denote the derivatives of the Airy functions. Equation \eqref{eq:BetaSq} gives the number of created particles of the mode $k$ up to the time $\eta$. The particle creation rate is obtained by taking the derivative of Eq. \eqref{eq:BetaSq}. In this case we obtain
\begin{align}\label{eq:BetaDeriv}\nonumber
\frac{d|\beta_k|^2}{d\eta}= &\frac{\pi z^2\eta}{4k}\Big[\Ai\Big( \frac{-k^2-z^2\eta}{z^{4/3}}\Big)\Ai'\Big( \frac{-k^2-z^2\eta}{z^{4/3}}\Big)\\
&+\Bi\Big( \frac{-k^2-z^2\eta}{z^{4/3}}\Big)\Bi'\Big( \frac{-k^2-z^2\eta}{z^{4/3}}\Big) \Big].
\end{align} 

\begin{figure}[H]
\includegraphics[width=1.0\columnwidth]{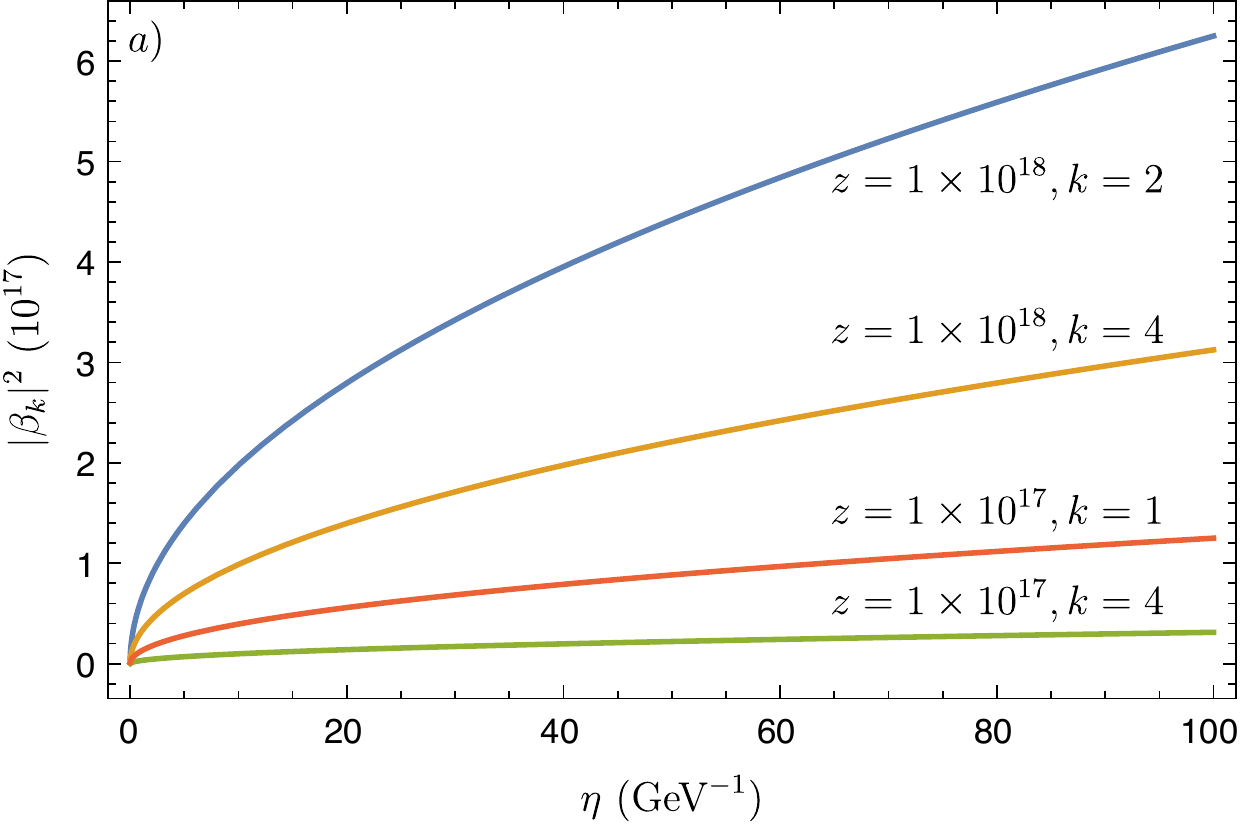}
\includegraphics[width=1.0\columnwidth]{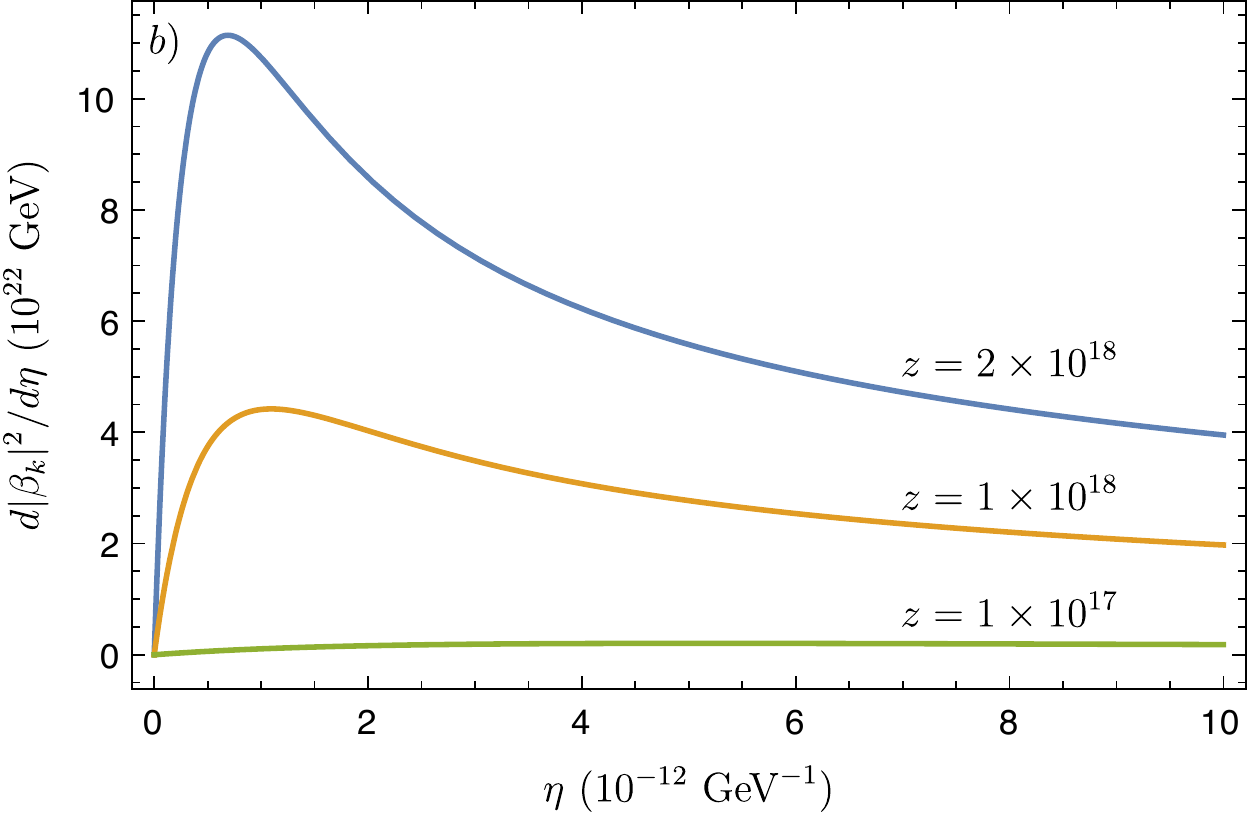}
\caption{Particle production for different values of $z$ and momenta $k$: $a)$ phase space number density, and $b)$ particle creation rate for $k=1$. The units of $z$ and $k$ are $\mathrm{GeV}^{3/2}$ and $\mathrm{GeV}$, respectively.}
\label{Fig:B_k}
\end{figure}

The behavior of particle production in time can be examined by taking a look at the graph of Eq. \eqref{eq:BetaSq} and Eq. \eqref{eq:BetaDeriv} for different values of $z$ and $k$. Figure \ref{Fig:B_k} gives the produced particle number and the particle creation rate as a function of the conformal time $\eta$ and illustrates how these depend on the parameter $z$ and momentum $k$.

\noindent Figure \ref{Fig:B_k}a illustrates that particle production is strongest for small momenta $k$ for all values of $z$. On the other hand, Fig. \ref{Fig:B_k}b reveals that particle creation is most intense when the parameter $z$ is large. Thus, for a fixed expansion parameter $b$ particle creation is most effective for a massive field, with large $z= mb$, producing essentially particles having a small momentum, i.e., being non-relativistic.

\subsection{Particle Energy and Thermalization}
\begin{figure}[H]
\includegraphics[width=1.0\columnwidth]{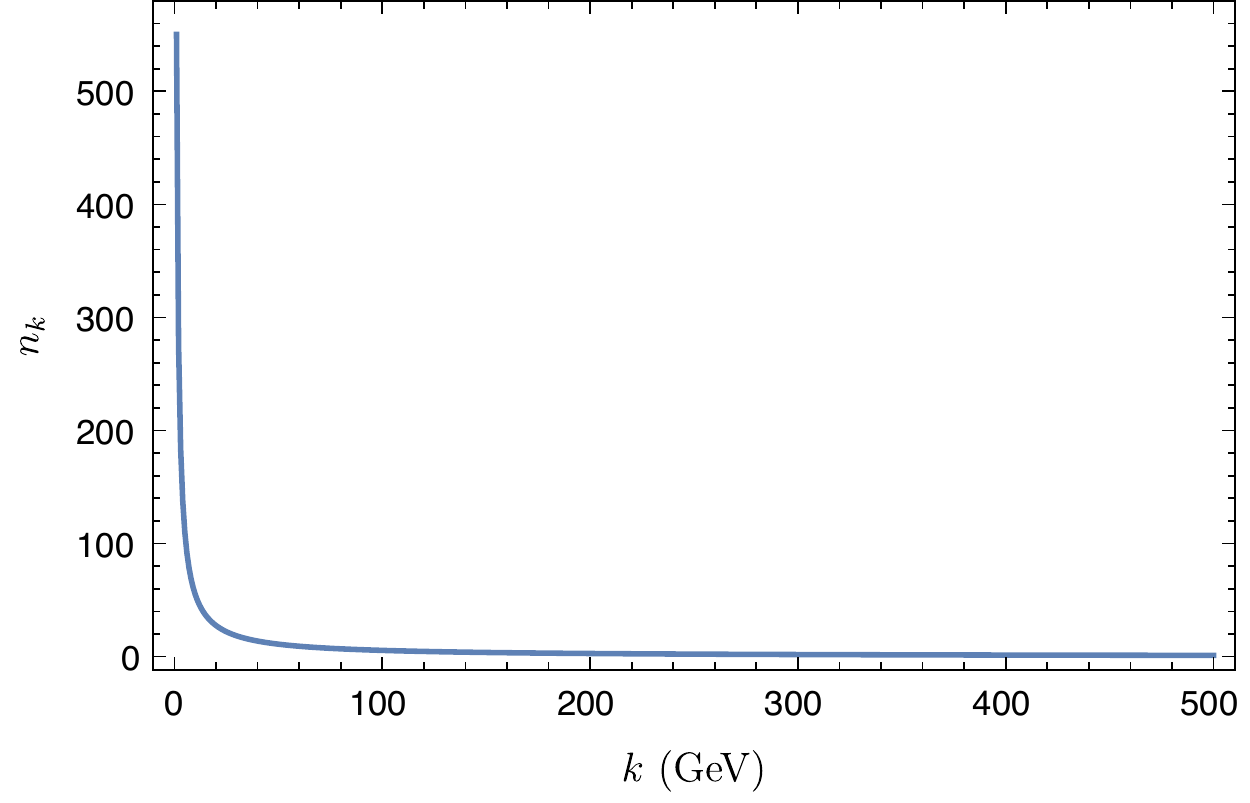} 
\caption{Particle number density $n_k$ per mode $k$ for $z=10^{20}\ \mathrm{GeV}^{3/2},\eta_0=10^{-20}\ \mathrm{GeV}^{-1}$ and $\eta=10^{-7}\ \mathrm{GeV}^{-1}$.}
\label{Fig:N_k}
\end{figure}

\begin{figure}[H]
\includegraphics[width=1.0\columnwidth]{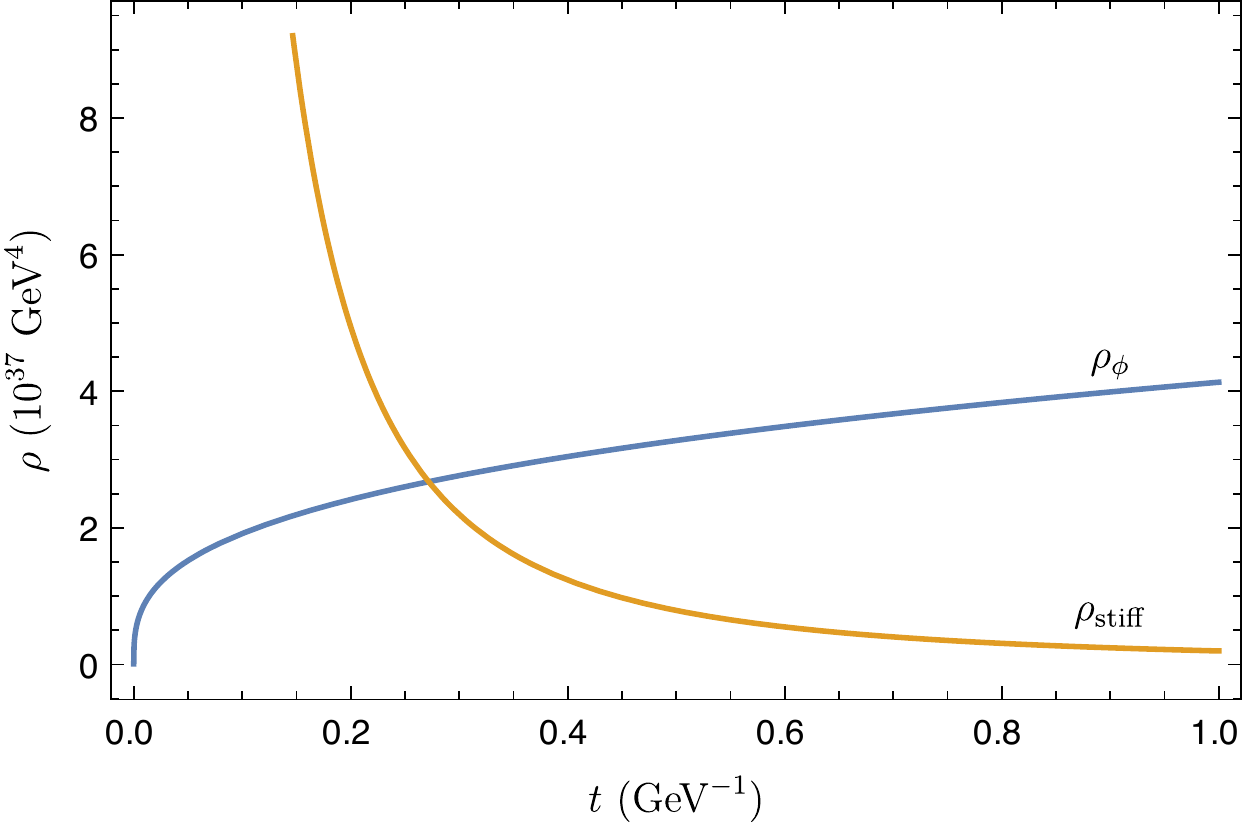} 
\caption{Energy densities $\rho_{\rm{stiff}}$ and $\rho_{\phi}$, when $m=10^8\  \rm{GeV}$ and $b=10^{10}\  \rm{GeV}^{1/2}$. Here $t_{\rm{eq}}\approx 0.27\ \mathrm{GeV}^{-1}$ corresponding to $\eta\approx 1.18\times 10^{-7}\ \mathrm{GeV}^{-1}$.}
\label{Fig:leikkaus}
\end{figure}

The energy density $\rho_{\phi}$ of the scalar particles created from some initial time $\eta_0$ up to time $\eta$ is given by 
\begin{align}\label{eq:ETCP_Int}
\rho_{\phi}(\eta)=\frac{1}{2\pi^2}\int_{\eta_0}^{\eta}d\tilde{\eta}\int_0^\infty dk\,k^2 \omega_k(\tilde{\eta})\Big[\frac{C(\eta_0)}{C(\tilde{\eta})} \Big]^{3/2} \frac{d|\beta_k|^2}{d\tilde{\eta}},
 \end{align}
where $\omega_k(\tilde{\eta})=(k^2+z^2\tilde{\eta})^{1/2}$. The factor $[C(\eta_0)/C(\tilde{\eta})]^{3/2}$ accounts for the dilution of the created particles caused by the expansion of space. In order to perform the integral \eqref{eq:ETCP_Int} it is necessary to make a simplification: we approximate $\omega_k(\tilde{\eta})\approx z\tilde{\eta}^{1/2}$ since the particle production rate is peaked to non-relativistic modes. The validity of the approximation can be justified by looking at the particle number density per mode
\begin{align}
n_k=\int_{\eta_0}^\eta d\tilde{\eta}\, \Big[\frac{C(\eta_0)}{C(\tilde{\eta})} \Big]^{3/2} \frac{d|\beta_k|^2}{d\tilde{\eta}},
\end{align}
taking into account the dilution. From  Fig. \ref{Fig:N_k} one sees that effectively only the low-momenta scalars are present in the universe if the scalar mass is more than about 100 GeV.
\noindent Performing the integration in Eq. \eqref{eq:ETCP_Int} and changing the variable $\eta$ to the standard coordinate time
\begin{align}
t=\frac{2}{3}b\eta^{3/2},
\end{align}
we obtain the energy density $\rho_{\phi}$ for the created particles as a function of coordinate time as
\begin{widetext}
\begin{align}
\label{eq:Rho_cp}\nonumber
\rho_{\phi}(t)=&\frac{3m^{11/3}b^{8/3}t_0}{64\pi}
\Big\{\!-\!\Big(\frac{3mt_0}{2}\Big)^{2/3}\!\Ai\Big[\!-\!\Big(\frac{3mt_0}{2}\Big)^{2/3}\Big]^2
\!+\!  \Big(\frac{3mt}{2}\Big)^{2/3}\!\Ai\Big[\!-\!\Big(\frac{3mt}{2}\Big)^{2/3}\Big]^2
\!-\!  \Ai'\Big[\!-\!\Big(\frac{3mt_0}{2}\Big)^{2/3}\Big]^2\\\nonumber
&\!+\!   \Ai'\Big[\!-\!\Big(\frac{3mt}{2}\Big)^{2/3}\Big]^2
\!-\!  \Big(\frac{3mt_0}{2}\Big)^{2/3}\Bi\Big[\!-\!\Big(\frac{3mt_0}{2}\Big)^{2/3}\Big]^2
\!+\!\Big(\frac{3mt}{2}\Big)^{2/3}\Bi\Big[\!-\!\Big(\frac{3mt}{2}\Big)^{2/3}\Big]^2
\!-\!\Bi'\Big[\!-\!\Big(\frac{3mt_0}{2}\Big)^{2/3}\Big]^2\\
&\!+\!\Bi'\Big[\!-\!\Big(\frac{3mt}{2}\Big)^{2/3}\Big]^2
\Big\},
\end{align}
\end{widetext}
where $t_0$ is the initial coordinate time and the parameters $m$ and $b$ have been reinserted.
Taking the limit $t\to\infty$ in Eq. \eqref{eq:Rho_cp} we notice that the energy density $\rho_\phi$ diverges meaning that the creation rate of the particles surpasses the dilution effect from the expansion of space. Hence, at some time $t_{\rm{eq}}$ the energy density of the scalar particles and the energy density of the stiff matter will be equal and the stiff matter dominated era ends. At the same time the stiff matter induced particle production ends. To evaluate this time we use $\rho_{\rm{stiff}}(t_{\rm{eq}})= \rho_{\phi}(t_{\rm{eq}})$, where
\begin{align}\label{eq:Rho_stiff}
\rho_{\rm{stiff}}(t)=\frac{1}{24\pi G t^2},
\end{align}
which can be solved numerically by fixing the values of the parameters $m,b$ and $t_0$. 
A natural choice for $t_0$ is the Planck time $t_{\rm{pl}}$ since for times under $t_{\rm{pl}}$ quantum gravitational effects cannot presumably be neglected. By fixing the initial time to be $t_{\rm{pl}}\approx 8.19\times 10^{-20}\ \mathrm{GeV}^{-1}$, we can numerically solve the time $t_{\rm{eq}}$ for different values of $m$ and $b$. Figure \ref{Fig:leikkaus} shows graphs of $\rho_{\rm{stiff}}(t)$ and $\rho_{\phi}(t)$.

To proceed, we assume that the created scalar particles decay fast to ordinary relativistic particles.
When thermalized, their maximal possible temperature is $T_{\rm{max}}$ which is the upper bound obtained from Eq. \eqref{eq:T_eq}.
Table \ref{Tab:1} shows the values of temperature $T_{\rm{max}}$ with some values of parameters $b$ and $m$. 
 
\begin{ruledtabular}
\begin{table}[H]
\caption{Temperature $T_{\rm{max}}$ in units of $\mathrm{GeV}$ for different values of $b$ and $m$ in units of $\rm{GeV}^{1/2}$ and $\rm{GeV}$ respectively.}
\begin{tabular}{l|lll}
\diagbox[width=1.8cm, height=0.5cm]{$b$}{$m$} & $1.00\times 10^{6}$ & $1.00\times 10^{8}$ & $1.00\times 10^{10}$ \\
\midrule
$1.00\times 10^{0}$		   & $8.48\times 10^{1}$	  & $4.39\times 10^{3}$	     & $2.27\times 10^{5}$		\\
$1.00\times 10^{10}$  & $4.39\times 10^{7}$	  & $2.27\times 10^{9}$	     & $1.18\times 10^{11}$		\\
$1.00 \times 10^{26}$  & $1.54\times 10^{16}$  & $6.41\times 10^{17}$	     & $4.14\times 10^{18}$	    	\\
\end{tabular}
\label{Tab:1}
\end{table}
\end{ruledtabular}

It can be seen, that the equilibrium temperature $T_{\rm{max}}$ is not only dependent on the mass of a given field, but also from the parameter $b$ controlling the expansion rate of the universe. By increasing this parameter up to about $10^{26}$, high temperatures can be reached even for light scalars.
Moreover, the larger $b$ is the faster the particles reach equilibrium. When $b=10^{26}\  \rm{GeV}^{1/2}$ the time it takes to reach equilibrium is about $10^5\ t_{\rm{pl}}$ for a scalar field with mass $10^{6}\  \rm{GeV}$.
However, if the expansion rate parameter $b$ is too small and the scalar mass is relatively light, thermalized particles do not even reach temperatures above $100\  \mathrm{GeV}$ to guarantee the possibility for electroweak baryogenesis; preferably the temperature should be at least $\sim 1\  \mathrm{TeV}$.

\section{Discussion}\label{sec:IV}
We have provided a model for studying the gravitational particle creation in a stiff matter dominated universe. The results show, that the particle creation is dependent on the mass $m$ and the momenta $k$ of a given scalar field and the expansion parameter $b$ in such a way that for a fixed value of $b$ it is most effective for a very massive field. 

We have also calculated an upper bound for the equilibrium temperature $T_{\rm{th}}$ supposing rapid decay of the scalar particles to ordinary matter. The obtained maximal equilibrium temperature depends on the parameters such that for large $m$ and $b$ very high temperatures are reached. For different values of the pair $(m,b)$, the equilibrium temperature ranges anywhere from few $\mathrm{MeV}$ up to about $10^{18}\ \mathrm{GeV}$. Realistically the temperature should reach values above $\sim 1\ \rm{TeV}$ to ensure a possibility for baryogenesis. This means that for a small value of the expansion parameter $b$ the scalar field must be very massive.
On the other hand, if $b$ is raised to higher values, even the lightest scalars attain temperatures way above $1\ \rm{TeV}$. Since there is a region of values of the parameters where the temperature $T_{\rm{max}}$ obtains values above the threshold of $\sim 1\ \rm{TeV}$, the model can be considered a viable one.

The assumptions we have made regarding the attained values of $T_{\rm{max}}$ require some discussion. First of all, given that we have not taken into account the decay rate of the scalar particles, the actual equilibrium temperature might be much lower than the temperature obtained with the made assumptions. Depending on how fast the scalars decay, the universe might be dominated by ordinary matter for some time before radiation domination is achieved. Also, after the equilibrium time $t_{\rm{eq}}$ there is still some stiff matter left. However, since it scales as $\rho_{\rm{stiff}}(t)\propto a(t)^{-6}$ it is quickly diluted away. Secondly, by fixing the initial time $t_0$ in the scalar particle energy density to be the Planck time $t_{\rm{pl}}$, we have assumed that the universe emerged as stiff matter dominated. Moving the initial time forward in time has consequences on the energy density and hence on the equilibrium temperature. If applied to inflationary scenarios, the initial time could be fixed at the reheating time. In this case the temperature needs to be high enough to be physically sound. Technically there is no problem in reaching the desired temperatures however, since increasing the parameter $b$ these can easily be reached. However, realistic $b$ parameter values remain to be determined.

Gravitational particle creation during a stiff matter dominated era has not been studied exhaustively and the few instances where it has appeared deal with inflationary situations \citep{deHaro:2016a,deHaro:2016b}. In these studies, the particle creation process is very different, since it takes place during an abrupt phase transition at the end of inflation to the stiff matter era. In our model the particles are produced during the stiff matter era also taking into account the expansion of the universe. 
Although there are similarities between the models, the results are not directly comparable to works of de Haro et al. \citep{deHaro:2016a,deHaro:2016b}.

The model we have presented in this paper offers a novel approach to gravitational particle creation in a stiff matter dominated universe opening up new and interesting aspects regarding particle creation. Along the way we made some simplifying approximations, which can be considered in greater detail. 
In particular inclusion of the finite decay time of the scalar particles, the finite thermalization time of the decay products as well as possibility to have e.g., post-inflatory stiff matter era could be relevant direction developments. These are considerations which we leave to future research.


\begin{thebibliography}{99}

\bibitem{Birrell_Davies}
 N.D. Birrell and P.C.W. Davies,
 \emph{Quantum Field Theory in Curved Space},
 (Cambridge University Press, Cambridge, 1982).
 
\bibitem{Parker:1968}
 L. Parker,
 \newblock{Phys. Rev. Lett. \textbf{21}, 562 (1968)}.  
 
\bibitem{Parker:1969}
 L. Parker,
 \newblock{Phys. Rev. \textbf{183}, 1057 (1969)}. 
 
\bibitem{Zeldovich_Starobinsky:1972}
 Y.B. Zel'dovich and A. A. Starobinsky,
 \newblock{Zh. Eksp. Teor. Fiz. \textbf{61}, 2161 (1971) [Sov. Phys. JETP \textbf{34}, 1159 (1972)]}.
 
\bibitem{Zeldovich_Starobinsky:1977}
 Y.B. Zel'dovich and A. A. Starobinsky,
 \newblock{Pis'ma Zh. Eksp. Teor. Fiz. \textbf{26}, 373 (1977) [JETP Lett. \textbf{26}, 252 (1977)]}. 
 
\bibitem{Hawking:1975}
 S.W. Hawking,
 \newblock{Commun. Math. Phys. \textbf{43}, 199 (1975)}.
 
\bibitem{Haro_Elizalde:2015}
 J. Haro and E. Elizalde,
\newblock{J. Cosmol. Astropart. Phys. \textbf{2015}, 028 (2015)}.
 
\bibitem{Tavakoli_Fabris:2015}
 Y. Tavakoli and J. Fabris,
\newblock{Int. J. Mod. Phys. D \textbf{24}, 1550062 (2015)}. 
 
\bibitem{Quintin:2014}
 J. Quintin, Y-F. Cai, and R. Brandenberger,
 \newblock{Phys. Rev. D \textbf{90}, 063507 (2014)}.  
 
\bibitem{Pereira:2014}
     S. Pereira and F. Holanda,
 \newblock{Gen. Relativ. Gravit. \textbf{46}, 1699 (2014)}.   


\bibitem{Fedderke:2015}
 M. Fedderke, E. Kolb, and M. Wyman,
 \newblock{Phys. Rev. D \textbf{91}, 063505 (2015)}.    
 
\bibitem{FRW} The so-called concordance model is described in numerous textbooks, see {\em e.g.,} S. Dodelson, {\em Modern Cosmology},
(Academic Press, 2008).
 
 
\bibitem{Zeldovich:1972}
Y.B. Zel'dovich,
\newblock{Mon. Not. R. Astr. Soc. \textbf{160}, 1P (1972)}.

\bibitem{Barrow:1978}
J.D. Barrow,
\newblock{Nature (London), \textbf{272}, 211 (1978)}.

\bibitem{Kamionkowski_Turner:1990}
M. Kamionkowski and M. Turner, 
\newblock{Phys. Rev. D \textbf{42}, 3310 (1990)}.

\bibitem{Dutta_Scherrer:2010}
S. Dutta and R. Scherrer,
\newblock{Phys. Rev. D \textbf{82}, 083501 (2010)}.

\bibitem{Joyce_Prokopec:1998}
M. Joyce and T. Prokopec,
\newblock{Phys. Rev. D \textbf{57}, 6022  (1998)}.

\bibitem{Oliveira:2011}
G. Oliveira-Neto, G. Monerat, E. Corr\^ea Silva, C. Neves and L. Ferreira-Filho,
\newblock{Int. J. Mod. Phys. Conf. Ser. \textbf{03}, 254 (2011)}.

\bibitem{Chavanis:2015}
P.H. Chavanis,
\newblock{Phys. Rev. D \textbf{92}, 103004, (2015)}.


\bibitem{Guth:1981}
A. Guth,
\newblock{Phys. Rev. D \textbf{23}, 347, (1981)}.

\bibitem{Linde:1982}
A. Linde,
\newblock{Phys. Lett. B \textbf{108}, 389, (1982)}.

\bibitem{PLANCK}
See {\em e.g.,} Planck Collaboration, R. Adam {\em et al.}, Astronomy and Astrophysics \textbf{594}, A13 (2016), and
 Planck Collaboration, R. Adam {\em et al.}, Astronomy and Astrophysics \textbf{594}, A20 (2016).


\bibitem{deHaro:2016a}
J. de Haro, J. Amor\'os and S. Pan,
\newblock{Phys. Rev. D \textbf{93}, 084018, (2016)}.

\bibitem{deHaro:2016b}
J. de Haro and E. Elizalde
\newblock{Gen. Relativ. Gravit. \textbf{48}, 77, (2016)}.  

\bibitem{Chun:2009}
E. Chun, S. Scopel and I. Zaballa,
\newblock{J. Cosmol. Astropart. Phys. \textbf{2009}, 022 (2009)}.  

\bibitem{Abramowitz_Stegun}
 M. Abramowitz and I. Stegun,
 \emph{Handbook of Mathematical Functions 9th edition},
 (Dover Publications, New York, 1972). 

  

\end{thebibliography}
\end{document}